\documentclass[aps,pra,twocolumn,showpacs,floatfix]{revtex4-2}
\usepackage{graphicx}
\usepackage{amsmath}
\usepackage{amssymb}
\usepackage{color}
\newcommand{\otwo}[0]{$\mathrm{O_2}$}
\newcommand{\ntwo}[0]{$\mathrm{N_2}$}
\newcommand{\cotwo}[0]{$\mathrm{CO_2}$}
\newcommand{\urad}[0]{$\mu\text{rad}$}
\newcommand{\lbrot}[0]{LB$_\text{rot}$}
\newcommand{\lbsys}[0]{LB$_\text{sys}$}

\begin{document}

\title{Observation of mechanical Faraday effect in gas media}

\author{Alexander~A.~Milner$^{1}$, Uri~Steinitz$^{2,3}$,  Ilya~Sh.~Averbukh$^{3}$, and Valery~ Milner$^{1}$}

\affiliation{$^{1}$Department of  Physics \& Astronomy, The University of British Columbia, Vancouver, Canada \\
$^{2}$Soreq Nuclear Research Centre, Yavne, Israel \\
$^{3}$AMOS and Department of Chemical and Biological Physics, The Weizmann Institute of Science, Rehovot, Israel}

\date{\today}

\begin{abstract}
We report the experimental observation of the rotation of the polarization plane of light propagating in a gas of fast-spinning molecules (molecular super-rotors). In the observed effect, related to Fermi's prediction of ``polarization drag'' by a rotating medium, the vector of linear polarization tilts in the direction of molecular rotation due to the rotation-induced difference in the refractive indices for the left and right circularly polarized components. We use an optical centrifuge to bring the molecules in a gas sample to ultrafast unidirectional rotation and measure the polarization drag angles of the order of 0.2 milliradians in a number of gases under ambient conditions. We demonstrate an all-optical control of the drag magnitude and direction, and investigate the robustness of the mechanical Faraday effect with respect to molecular collisions.
\end{abstract}
\maketitle

Faraday effect is typically associated with the rotation of the polarization plane of light traveling through the medium subject to external magnetic field. Magnetic field introduces a nonzero difference between the refractive indices for two orthogonal circularly polarized components. This difference results in the changing relative phase between the circular components and leads to the rotation of the field polarization. Faraday effect also has a mechanical analogue: polarization rotation due to the mechanical rotation of the medium (known as \textit{polarization drag}). In this case, the angular Doppler effect \cite{Garetz1981} causes the opposite frequency shifts for the two circular polarization components in the rotating reference frame, and an accumulated phase lag between them due to the optical dispersion. In addition to the dispersive component, polarization drag in a rigidly rotating medium also has a relativistic component, first considered by Fermi in 1923 \cite{Fermi1923} and analysed further by others \cite{Player1976,Evans1992,Nienhuis1992}.

Polarization drag in a rotating solid was first experimentally observed by Jones in a glass sample which was spun to the rotational frequency of up to 140~Hz \cite{Jones1976}. That experiment has been a major tour de force, requiring sensitivity to polarization rotation below 0.1~$\mu $rad for detecting the drag angles on the scale of just a few microradians. A related effect was recently reported in \cite{Franke2011}, where the drag effect was enhanced by many orders of magnitude in a spinning ruby rod due to the resonant interaction resulting in the extremely slow group velocity of the probe light.

No observation of polarization drag in a gas medium has been reported to date, despite the attention it has received in theoretical works \cite{Nienhuis1992} and its potentially important role in astrophysical applications \cite{Gueroult2019}. In comparison to solids, the much lower density of gases presents a major challenge for studying this effect in the laboratory. However, a recent proposal for observing mechanical Faraday effect in gases suggested to use the fast unidirectional rotation of individual molecules as a way of compensating for the low sample density \cite{Steinitz2020}.

\begin{figure*}[t]
    \includegraphics[width=0.85\textwidth]{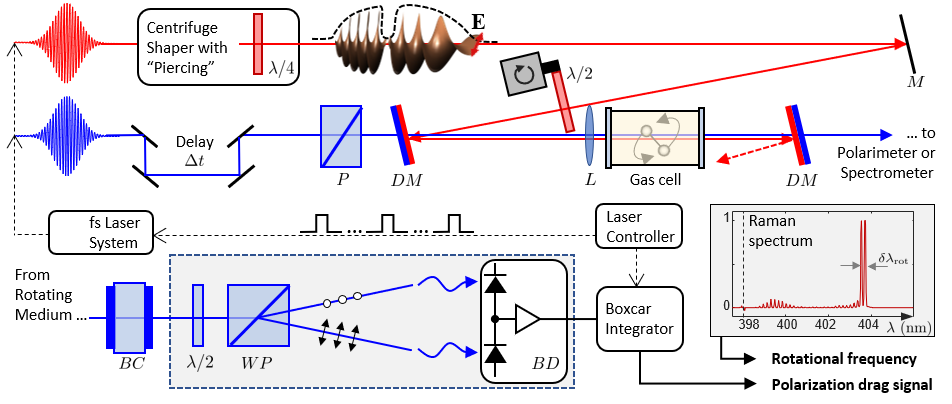}
    \caption{Scheme of the experimental setup. Top: femtosecond pulses with the central wavelength of 800~nm (upper, red) and 400~nm (lower, blue) are used for creating the centrifuge and the probe pulses, respectively. The pulses are shaped, delayed with respect to one another, combined in a collinear geometry and focused in a gas cell. Bottom: after passing through the gas sample, probe pulses are filtered out from the centrifuge light and sent to the time-gated polarization analyzer, implemented with a boxcar integrator triggered at the time of arrival of probe pulses at the detector. $\lambda/4 (\lambda/2)$: quarter(half)-wave plate, $P$: polarizer, $M$: metallic mirror, $DM$: dichroic dielectric mirror, $L$:lens, $BC$: Berek compensator, $WP$: Wollaston prism, $BD$: balanced detector. Alternatively, the probe pulses may be sent to a Raman spectrometer. The inset at the bottom right depicts an example of the Raman spectrum, representing a rotational wave packet of width $\delta \lambda_\text{rot}$ in oxygen. The dashed line at 398~nm labels the central frequency of the probe pulses.}
    \label{Fig-setup}
\end{figure*}
Here, we report on the first experimental demonstration of the mechanical Faraday effect in gases. We optically bring molecules to the state of ultrafast rotation (known as a super-rotor state, SR) by means of a laser tool known as an optical centrifuge \cite{Karczmarek1999,Villeneuve2000,Yuan2011,Korobenko2014}, and measure the induced change in the polarization of a weak probe pulse following the centrifuge excitation.

We observe polarization drag angles of the order of $0.2$~mrad for a propagation distance of 1~mm in various gases under ambient conditions (\otwo{}, \ntwo{}, \cotwo{} and ambient air) when the maximum rotational frequency of molecules reaches $\sim6$~THz. This agrees well with the theoretical expectation of about 1~mrad of polarization rotation per 1~THz of average rotational frequency in a gas medium under similar conditions \cite{Steinitz2020}, since only a few percent of the molecules are typically trapped in the centrifuge and reach the high degree of rotational excitation \cite{Milner2015c}. Expressed in the units of drag angle per sample's density and propagation length, the observed rotary power is orders of magnitude higher than previously observed in solids \cite{Jones1976, Franke2011}, as the laser-induced ultrafast rotation more than makes up for the low gas density. The effect is shown to last on a nanosecond time scale, extending well beyond the centrifuge pulse, until tens of collisions finally cause the molecular super-rotors to spin down \cite{Khodorkovsky2015, Milner2015c, Steinitz2016}. We also demonstrate that the angle of polarization rotation can be optically controlled by changing the rotational frequency of the centrifuged molecules.

An optical centrifuge is a laser pulse, whose linear polarization rotates at an accelerating rate \cite{Karczmarek1999, Villeneuve2000}. Our setup for producing the centrifuge has been described in a recent review \cite{MacPhail2020}. Briefly, we split the spectrum of broadband laser pulses from a Ti:S amplifier (10 mJ, 35 fs, repetition rate 1~KHz, central wavelength 795~nm) in two equal parts using a Fourier pulse shaper. The two equal-amplitude beams are frequency chirped with opposite chirp signs and have opposite circular polarizations. When combined together, optical interference of these laser fields results in the rotation of the polarization vector with an instantaneous frequency growing linearly in time from 0 to 10~THz over the course of about 100~ps. Molecules interact with the centrifuge field via the induced electric dipole moment and, if the interaction potential exceeds their thermal energy, follow the accelerated rotation of the centrifuge.

The centrifuge pulses were focused in a cell filled with a gas of interest at room temperature and atmospheric pressure, as schematically illustrated at the top of Fig.~\ref{Fig-setup}. The focusing lens $L$ with a focal length of 10~cm provided the length of the centrifuged region of about 1~mm and a peak intensity of up to $5\times 10^{12}$~W/cm$^{2}$. Measuring both the rotational frequency of SRs and the polarization drag angle was done with short probe pulses (pulse lengths of $\sim3$~ps) delayed with respect to the centrifuge. The probe pulses were derived from the same laser system, spectrally narrowed down to the bandwidth of $0.1$~nm, and frequency doubled for the ease of separating them from the excitation light. Care was taken to make the probe focal spot smaller than the corresponding size of the centrifuge beam to minimize the effects of pointing instability and density gradients, discussed later in the text.

To characterize molecular rotation, both in terms of its frequency and sense, we used coherent Raman spectroscopy. Coherent scattering of narrowband probe pulses from rotating light molecules, such as \otwo{} or \ntwo{}, resulted in Raman spectra with well-resolved peaks, corresponding to individual rotational quantum states (see inset in Fig.~\ref{Fig-setup}). The magnitude of the Raman shift was translated to the rotational frequency, whereas its sign indicated the sense of rotation with respect to the circular probe polarization \cite{Korech2013, Korobenko2014a}. The shot-to-shot stability of the Raman spectra also served as a sensitive indicator of whether the detrimental strong-field effects, e.g. ionization and filamentation, were properly avoided and did not introduce noise to our sensitive polarization drag detection.

A very sensitive method of detecting small degree of polarization rotation is based on an optical configuration depicted inside the dashed gray rectangle at the bottom of Fig.~\ref{Fig-setup}. A half-wave plate is used to align the probe polarization at 45 degrees with respect to the axes of a Wollaston prism. This equalizes the intensity of light in both arms of a differential balanced detector resulting in a zero signal. As soon as probe polarization undergoes rotation in the sample medium, the balance shifts towards one of the photo-diodes yielding a non-zero signal, whose sign indicates the direction of rotation. Calibration is achieved via the rotation of the input polarization by a known amount using the same half-wave plate. This basic scheme, utilized in the pioneering experiment of R.~V.~Jones in 1976 was also implemented in our setup.

Because of the short lifetime of molecular rotation under ambient conditions, we used picosecond probe pulses (same pulses as in the Raman scattering experiment, but linearly polarized) to measure the polarization rotation angle at a given time delay. Signals from the amplified balanced detectors were gated around the arrival time of the probe pulses. To eliminate systematic errors due to the drifts in pulse energy, we alternated between clockwise (CW) and counter-clockwise (CCW) centrifuge handedness by applying a 90~degree rotation of the quarter-wave plate in the centrifuge shaper. Since the induced mechanical Faraday rotation is expected to follow the direction of the centrifuge, we calculated the polarization drag angle as the half-difference between the CW and CCW signals, scaled according to the calibration constant, obtained with the blocked centrifuge beam. To minimize the effect of dichroic dielectric mirrors on beams' polarization, those mirrors were set as close to normal incidence as possible.

In all of our previous studies, the rotational frequency of the super-rotors was controlled by truncating the centrifuge pulse in time, which resulted in an early termination of the accelerated molecular rotation and correspondingly lower frequency. This method involves significant changes of the total pulse energy by up to 50\%. Given the high centrifuge intensities, big variations in the pulse energy can lead to changes in the local heating of optical elements, potentially affecting their birefringence and modifying the polarization of the probe pulses passing through them. To eliminate any effects of the pulse energy on the sensitive polarimetry setup, we developed a new technique of centrifuge ``piercing'' \cite{Amani2021}. As schematically illustrated in Fig.~\ref{Fig-setup}, a short 2~ps notch was introduced in the field envelope of an optical centrifuge by means of a spectral filter in the centrifuge shaper. The notch interrupts the accelerated rotation of molecules, causing them to stop following the centrifuge. Controlling the rotational frequency was executed by moving the notch position in time, which was accompanied by less than 2\% variation in pulse energy.

Detection of the polarization drag in both media, i.e. a rotating piece of glass and a gas of rotating molecules, is susceptible to the same main source of systematic errors - a \textit{linear} birefringence (LB), which is created in the sample due to its forced rotation in addition to the \textit{circular} birefringence responsible for the mechanical Faraday rotation. Similarly to the latter, LB also results in the rotation of the polarization axes and may completely overwhelm and mask the more subtle drag effect. In the rotating glass experiment, LB  stems from the mechanical and thermal stress gradients, which were meticulously suppressed by means of the proper choice of material and careful mechanical design \cite{Jones1976}.

Despite the lack of mechanically moving parts in the case of a gas sample, suppressing the undesired linear birefringence is equally important and challenging. Here, LB stems from the anisotropic spatial distribution of the molecular axes during the time of the laser-induced molecular rotation, known as molecular alignment \cite{Stapelfeldt2003, Ohshima2010, Fleischer2012}. This anisotropy results in a refractive index difference for the probe field polarized along and perpendicular to the alignment axis of the order of $n_\parallel-n_\perp \approx 10^{-5}$  \cite{Renard2004}. Linear birefringence of this magnitude (hereafter referred to as \lbrot{}) would yield a quarter-wave retardance for a probe wavelength of 400~nm over a distance of 1~cm, completely changing its polarization state. Hence, both the short-term molecular alignment appearing on the time scale of a rotational period, and the long-term alignment averaged over many rotational periods, may cause non-negligible changes of probe polarization (which in fact is often used to quantify the degree of alignment \cite{Renard2003}). This laser-induced linear birefringence presented a major obstacle towards the detection of the polarization drag effect in a gas of rotating molecules.

One could naively assume that the contribution of the described effect to the measured polarization drag should be rather insignificant. Indeed, unidirectional molecular rotation initiated with circularly symmetric field of an optical centrifuge should have no preferential linear axes. And even if a small linear anisotropy in the field distribution does exist (for instance, owing to a chromatic or polarization dispersion of optical elements) and produces a weak molecular alignment, one could envision an easy way of cancelling its effect out: since the polarization drag flips sign upon changing the direction of molecular rotation, whereas the effect of the linear birefringence might be expected to remain the same, subtracting the two signals for the clockwise and the counter-clockwise rotating sample should get rid of the \lbrot{} related artefact.

\begin{figure}[t]
    \includegraphics[width=0.99\columnwidth]{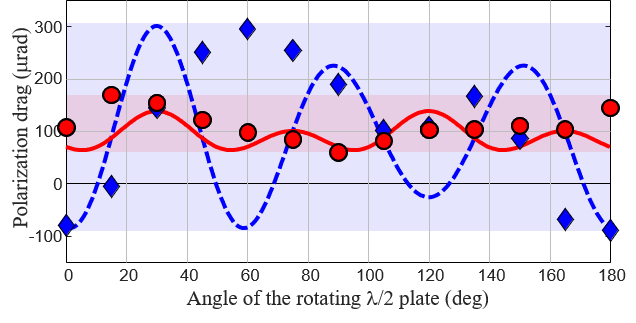}
    \caption{Markers: experimentally observed polarization drag signal with uncompensated (blue diamonds) and compensated (red circles) static linear birefringence of the optical system. Lines: calculated polarization rotation signal in the presence of polarization drag of 100~\urad{}, centrifuge-induced linear birefringence of $3.25\times 10^{-3}$~rad, and a combined static linear birefringence in the path of probe pulses of either $-0.44$~rad (dashed blue) or 0~rad (solid red). The birefringence units are angles of phase retardation between two orthogonal probe field polarizations.}
    \label{Fig-artefacts}
\end{figure}
Unfortunately, both assumptions -- the irrelevance of the weak induced molecular alignment and its independence on the direction of rotation, proved incorrect. To check for the presence of the centrifuge-induced linear birefringence, we inserted an additional half-wave plate in the path of the centrifuge beam and measured the polarization drag angle in a gas of oxygen SRs as a function of the wave plate's orientation. The results are shown with blue diamonds in Fig.~\ref{Fig-artefacts}. Hereafter, positive sign of the experimental signal means that the polarization was dragged in the direction of the centrifuge rotation. One can see in the figure that the drag signal spans a wide range of values (blue shaded area) from about $-100$~\urad{}, i.e. \textit{against} the direction of molecular rotation, to $+300$~\urad{} in that direction, around an average value of $\sim+100$~\urad{}. Since a perfect circularly isotropic centrifuge would simply flip its rotation sense \textit{regardless} of the angle of the inserted half-wave plate, the rather high sensitivity of the experimental signal to that angle (up to a sign reversal!) is indicative of the centrifuge-induced linear birefringence. We conclude that a slight linear anisotropy in the polarization structure of the centrifuge field may result in the small degree of molecular alignment, whose axis is rotated by the rotating half-wave plate and causes large spurious oscillations in the drag signal.

To confirm this model and establish the proper method for measuring the mechanical Faraday effect in a gas of rotating molecules, we calculated the expected signal, as obtained in our optical setup, using the standard methods of Jones calculus \cite{HechtBook}. In modeling the system, we made the following assumptions: (a) the magnitude of the polarization drag effect was set to $\sim100$~\urad{} in accordance with the average value detected experimentally; (b) we introduced a static (i.e. not related to the forced molecular rotation) combined linear birefringence of all optical components in the path of the probe beam (its exact value, hereafter referred to as \lbsys{}, will be explained later in the text); (c) the absolute value of the centrifuge-induced \lbrot{} was assumed to be the same for both the CW and CCW centrifuges, and used as a free parameter; and (d) since the centrifuge handedness is controlled by a 90 degree rotation of a quarter-wave plate, shown as part of the centrifuge shaper in Fig.~\ref{Fig-setup}, we assumed a 90 degree difference in the orientations of the CW and CCW centrifuge-induced birefringence. The result of the calculations, shown by the dashed blue curve in Fig.~\ref{Fig-artefacts}, demonstrates that the observed spread in the polarization drag angles may indeed be quite large despite the small values of \lbrot{}.

Fitting the calculated oscillatory shape (dashed blue curve) to the experimental values (blue diamonds) proved impossible with a single free parameter. Nevertheless, matching the peak-to-peak scatter of the measured drag signal to the numerically simulated one allowed us to estimate the magnitude of the centrifuge-induced \lbrot{}, which appears to be of the order of $3\times 10^{-3}$~rad in units of phase retardation. For an optical path of about 1~mm and a wavelength of 400~nm, this translates to the refractive index anisotropy $n_\parallel-n_\perp \approx 2\times 10^{-7}$, more than two orders of magnitude lower than what has been observed when molecular alignment was deliberately created with femtosecond pulses \cite{Renard2004}. Optical anisotropy of this minute strength is indeed not surprising in the case of creating unidirectional molecular rotation with strong polarized laser fields \footnote{Converting $\text{d}n=2\times 10^{-7}$ to the degree of molecular alignment yields $ \langle \cos^2\theta \rangle - 1/3 = 8\times 10^{-4}$, where $\theta$ is the angle between the molecular axis and a fixed direction in space, and $\langle ..\rangle$ indicates averaging over the molecular ensemble. In comparison, $ \langle \cos^2\theta \rangle - 1/3 \approx 0.3$ has been observed when molecules are intentionally aligned with femtosecond pulses under similar gas conditions\cite{Renard2004}.}, and can in fact be much higher if pairs of femtosecond pulses are used instead of the centrifuge \cite{Fleischer2009, Kitano2009, Zhdanovich2011, Karras2015}, resulting in much larger systematic errors.

The above analysis makes it clear that averaging the detected polarization drag signal over the angle of the half-wave plate, inserted in the path of an optical centrifuge, is key to quantifying its magnitude correctly. For that reason, we mounted the plate on a motorized rotational stage (a box labeled with ``$\circlearrowright$'' in Fig.~\ref{Fig-setup}), continuously rotating with the frequency of a few revolutions per second. At the same time, the disagreement between the calculated and experimentally measured values, which we attribute to the imperfections of the wave plate (chromaticity, non-parallelism, spatial non-uniformity, etc.), suggests that such averaging may not result in the true value of the polarization rotation angle. To further minimize the scatter in the experimental data, we notice that the numerically calculated peak-to-peak variation in the polarization drag signal decreases with the decreasing combined static birefringence of the optical system. Setting that value to zero resulted in the red solid line in Fig.~\ref{Fig-artefacts}. To lower \lbsys{} in the experiment, we inserted a variable wave plate (Berek compensator, $BC$ in Fig.~\ref{Fig-setup}) in the probe beam path. We then adjusted the parameters of Berek compensator to minimize the scatter in the polarization drag values (red circles), which brought it down to the numerically calculated range (red shaded area). Minimum spread was achieved with the compensator's birefringence set to 0.44~rad. We therefore used this value with an opposite sign in the simulations of the uncompensated system (blue shaded area). The same procedure of (i) eliminating the static linear optical anisotropy and (ii) averaging over the angles of the rotating wave plate, was later repeated in all our experiments and proved invaluable for the correct determination of the centrifuge-induced Faraday rotation in gas samples.

\begin{figure}[t]
    \includegraphics[width=0.99\columnwidth]{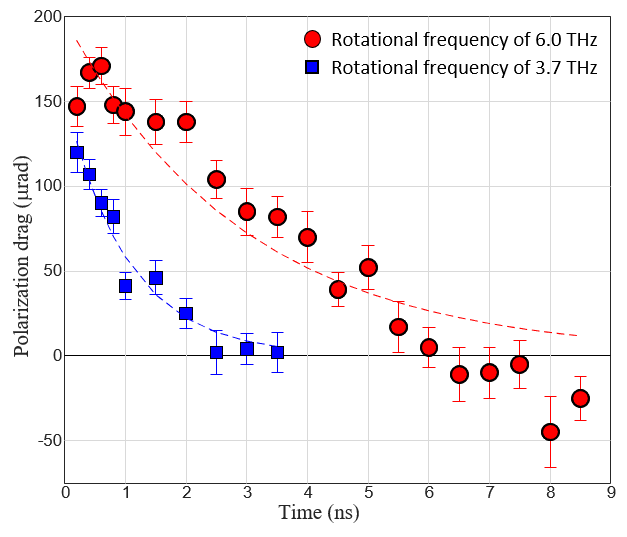}
    \caption{Experimentally measured decay of the polarization drag signal in time for two rotational frequencies of oxygen super-rotors: 6.0~THz (red circles) and 3.7~THz (blue squares). Dashed lines show the best fits to the exponential decay towards zero drag with time constants of 3.0~ns and 1.0~ns, respectively. Oxygen gas was kept at room temperature and pressure of 1~atm. Vertical error bars represent the standard error of the mean over $10^{5} - 10^{6}$ laser pulses.}
    \label{Fig-timedep}
\end{figure}
The developed experimental procedure enabled us to detect a polarization rotation angle of as low as $\sim10$\urad{}. The sensitivity reached previously in the experiments with rotating solids \cite{Jones1976} was about 300 times higher compared to our measurements, owing to a number of favorable factors, such as significantly longer effective sample length ($800$~mm of glass vs shorter than $1$~mm in the case of a centrifuged gas sample) and a practically infinite lifetime of the mechanical Faraday effect in a rotating solid vs a quick collisional decay of the centrifuge-induced molecular rotation on a ns time scale. This decay required short probe pulses with much lower stability, both in intensity and pointing, than the corresponding stability of a typical continuous wave laser source. The lower stability is particularly detrimental in the pump-probe geometry required here in comparison to a single continuous wave probe geometry in \cite{Jones1976}. However, in accord with the theoretical prediction \cite{Steinitz2020}, employing an optical centrifuge led to the drag angles as high as 180~\urad{} or 60 times higher than that observed in a rotating glass rod, making our resolution sufficient for detecting the rotation-induced polarization drag in gases, despite the lower density of gas media and the shorter effective optical length.

As super-rotors collide, their rotational frequency decreases and their rotational energy is transferred to thermal molecular motion. When the rotational frequency decreases, the mechanical Faraday effect also decays, as confirmed by our experimental results shown in Fig.~\ref{Fig-timedep}. Red circles correspond to the initial frequency of centrifuged oxygen molecules of 6~THz, as measured by means of coherent Raman scattering described earlier in the text. The non-exponential decay of the effect in time stems from two reasons. First, high rotational frequencies of SRs result in complex relaxation dynamics with a quiescent period followed by an abrupt broadening of the rotational distribution and its eventual collapse towards thermal equilibrium \cite{Steinitz2016, Khodorkovsky2015}.
\begin{figure}[t!]
    \includegraphics[width=0.99\columnwidth]{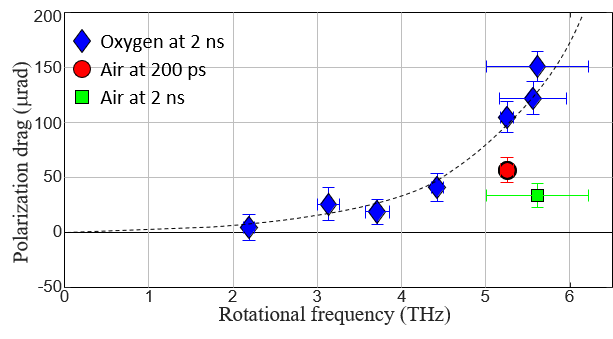}
    \caption{Experimentally measured dependence of the polarization drag signal on the initial rotational frequency of oxygen super-rotors 2~ns after their release from an optical centrifuge (blue diamonds). The dashed line is drawn to guide the eye. The red circle and the green square correspond to the drag signal in ambient air at 200~ps and 2~ns time delay after the centrifuge, respectively. The horizontal bars represent the width of the rotational wave packet, determined by Raman spectroscopy ($\delta \lambda_\text{rot}$ in the inset to Fig.~\ref{Fig-setup}), rather than an error in its central frequency.}
    \label{Fig-freqdep}
\end{figure}

Second, a crossover from the directional rotation to thermal dynamics is accompanied by the formation of ``density depression channels'' inside the gas of centrifuged molecules \cite{Steinitz2012,Jhajj2014, Lahav2014, Milner2015c, Steinitz2016}. Local density gradients may result in non-uniform spatial birefringence distributions. Hence, small changes in the position of the probe beam relative to the centrifuge during the delay scan, as well as due to the imperfect pointing stability, made this birefringence hard to compensate. As shown in our numerical analysis above, uncompensated residual \lbsys{} affects the magnitude of the detected signal, shifting it in an unpredictable manner. Corresponding fluctuations in the experimental data became especially pronounced after about 7~ns, when the depression channel starts dominating the propagation of probe pulses \cite{Milner2015c}. Note that the time constant of 3~ns is close to that found in our study of the crossover from rotational to thermal dynamics \cite{Milner2015c}.

In order to limit the influence of the rotation-translation energy transfer on the detected polarization drag signal, we lowered the rotational frequency of O$_{2}$ molecules from 6 to 3.7~THz (blue squares). This almost three-fold decrease of rotational energy, from 0.93~eV to 0.35~eV per molecule, yielded a better fit to an exponential decay, owing to the smaller local heating and weaker density gradients in the gas sample. As anticipated, the time constant shortened from 3 to 1~ns due to the faster collisional relaxation at lower rotation frequencies \cite{Hartmann2012, Milner2014a}. This signature of super-rotor dynamics demonstrates the potential of using the polarization drag measurement as a probe of the nonequilibrium gas kinetics.

A more detailed study of the mechanical Faraday effect as a function of the frequency of unidirectional molecular rotation is shown in Fig.~\ref{Fig-freqdep}. Measured in oxygen under the pressure of 1~atmosphere at a time delay of 2~ns after the centrifuge, the polarization rotation angle grows from zero to 150~\urad{} with increasing rotational frequency. The superlinear growth stems from the longer lifetime of faster super-rotors with respect to the slower ones. Hence, higher rotational frequencies increase the density of molecules undergoing unidirectional rotation at any given delay, in addition to the amount of polarization drag per molecule.

In summary, we report the first experimental observation of the mechanical Faraday effect (a.k.a. rotary polarization drag) in gas media. This effect, previously observed only in rotating solids, became accessible in a gas medium due to the high frequencies of molecular spinning ($\approx 6$~THz) induced and controlled by the optical centrifuge. The linear polarization of a weak probe light was ``dragged'' by the rotating oxygen molecules by an angle of up to $200$~\urad{} in the direction of their rotation. Polarization drag of comparable magnitude was also observed in \ntwo{} and \cotwo{} samples, as well as in ambient air (see Fig.~\ref{Fig-freqdep}). The dependence of the observed centrifuge-induced mechanical Faraday rotation on time and rotational frequency is in good qualitative agreement with the theoretically expected behaviour. The non-resonant nature of the effect and its robustness with respect to collisions means that it can be studied in other gases and under various conditions. Producing an equivalent magneto-optic Faraday effect in these gases would require an immense magnetic field of $\sim50~T$ \cite{ingersoll1954}.

The developed time-resolved polarimetry technique may be useful in probing the dynamics of super-rotors, such as their gyroscopic stability and their fast relaxation stage, as well as in the studies of polarization rotation in chiral molecules and its relation to their interaction with an optical centrifuge \cite{Tutunnikov2018, Milner2019}. Here, the revealed sensitivity of the apparent drag angle to minute amounts of linear birefringence, inevitably induced in gas samples by strong polarized laser pulses, must be considered carefully. The demonstrated method of examination and cancellation of birefringence-related errors should be useful in future experiments on light propagation in gas media interacting with strong polarized optical fields.

\section*{Acknowledgments}
We thank Ilia Tutunnikov for many useful discussions on the topic. This work was carried out under the auspices of the Canadian Center for Chirality Research on Origins and Separation (CHIROS). It was partially supported by the Israel Science Foundation (Grant No. 746/15). I.A. acknowledges support
as the Patricia Elman Bildner Professorial Chair, and thanks the UBC Department of Physics \& Astronomy for hospitality extended to him during his sabbatical stay.


%

\end{document}